\documentclass[aps, pra, twocolumn, floatfix, showpacs,superscriptaddress]{revtex4-1}
\usepackage{graphicx, amsfonts, amsmath, amssymb}
\usepackage[utf8]{inputenc}
\usepackage{adjustbox}
\usepackage{array}
\usepackage{color,empheq}
\usepackage{xcolor}
\usepackage{float}
\usepackage[colorlinks=true, citecolor=red, urlcolor=blue ]{hyperref}
\newcolumntype{L}[1]{>{\raggedright\let\newline\\\arraybackslash\hspace{0pt}}m{#1}}
\newcolumntype{C}[1]{>{\centering\let\newline\\\arraybackslash\hspace{0pt}}m{#1}}
\newcolumntype{R}[1]{>{\raggedleft\let\newline\\\arraybackslash\hspace{0pt}}m{#1}}
\usepackage{multirow}

\begin{document}
%opening
\title{
%Quantum heat engine using energy quantization and resources of degeneracy
Quantum heat engine based on level degeneracy
}
\author{George Thomas}
\email{georgethomas.iiserm@gmail.com}
\affiliation{Optics and Quantum Information Group, 
The Institute of Mathematical Sciences, HBNI,  CIT Campus, Taramani, Chennai 600113, India}
\affiliation{QTF centre of excellence, Department of Applied Physics, Aalto University School of Science, P.O. Box 13500, 00076 Aalto, Finland}
\author{Debmalya Das}
\email{debmalyadas@hri.res.in}
\affiliation{Harish-Chandra Research Institute, HBNI, Chhatnag Road, Jhunsi, Prayagraj (Allahabad) 211 019, India}
\author{Sibasish Ghosh}
\email{sibasish@imsc.res.in}
\affiliation{Optics and Quantum Information Group, 
The Institute of Mathematical Sciences, HBNI,  CIT Campus, Taramani, Chennai 600113, India}
%%%%%%%%%%%%%%%%%%%%%%%%%%%%%%%%%%%%%%%%%%%%%%%%%%%%%%%%%%%%%%%%%%%%%%%
\begin{abstract}
We study a quantum Stirling cycle which extracts work using quantized energy levels of a potential well. The
work and the efficiency of the engine depend on the length of the potential well, and the Carnot efficiency is
approached in a low temperature limiting case. We show that the lack of information about the position of the
particle inside the potential well can be converted into useful work without resorting to any measurement. In the
low temperature limit, we calculate the amount of work extractable from distinguishable particles, fermions, and
bosons.
\end{abstract}
\maketitle
%%%%%%%%%%%%%%%%%%%%%%%%%%%%%%%%%%%%%%%%%%%%%%%%%%%%%%%%%%%%%%%%%%%%%%
\section{Introduction}
The idea of Maxwell's demon occupies a central position in the understanding of thermodynamics and information.
It was introduced in a thought experiment that envisaged a situation in which there could be a possible violation 
of the second law of thermodynamics~\cite{Maxwell-Book,Leff-Book}. A classical analysis of Maxwell demon was 
first developed in the form of the Szilard engine~\cite{Szilard1929}. The Szilard engine consists of an enclosed
chamber
containing a gas molecule. A thin and massless  partition is inserted in the middle of the chamber~\cite{Maruyama2009}.
The 
demon measures the position of the molecule to the right or to the left of the partition and records it. Based
on the measurement, the demon then connects a mass to the partition on the same side as the molecule. Now by
absorbing heat from a hot bath, the gas can expand isothermally to occupy the original full volume of the chamber.
The partition, consequently, in pulling the mass, performs work of magnitude $k_B T\ln{2}$ where $T$ is the
temperature of the bath and $k_B$ is the Boltzmann constant. Superficially, it seems that the involvement
of the demon enables a Szilard engine, with a single gas molecule, to perform $k_B T\ln{2}$ amount of work, leading 
to a decrease of entropy of the heat bath, measuring $k_B \ln{2}$. This is impossible, according to the 
second law of thermodynamics, as a minimum of an equivalent increase of entropy is required in some part of the global
system. In~\cite{Szilard1929} it was suggested that an equivalent amount of work is required in the measurement of
the position of the gas molecule which saves the second law. However, it was not until~\cite{Landauer1961} that the 
work done in the erasure of information in the demon's memory was taken into consideration and the role played by 
measurement was refuted~\cite{Bennett1982, Landauer1961}. 
Landauer's erasure principle showed that minimum amount of increase in the entropy has to be $k_B \ln{2}$ 
for erasing  one-bit memory stored by the demon, establishing
an intriguing connection between information and thermodynamics
~\cite{Bennett1982, Landauer1961}. Further, Landauer's erasure principle has been experimentally established
using a single colloidal particle~\cite{Lutz2012}. An amount of work, which is nearly $k_B T\ln{2}$, has been experimentally
extracted from one bit of information,  using a single electron engine \cite{Koski2014}.

In the quantum version of the Szilard  engine, the insertion and the removal of the barrier constitutes a certain amount of work
and heat exchange, unlike in the classical case ~\cite{Kim2011}. Compared to the compression of the particle to the 
left (or right) side of the box, the 
insertion of the barrier needs a lower amount of work. This can be interpreted as follows.
In an insertion scenario, the position of the particle is unknown. One has to perform a measurement
after the insertion to
 determine the position of the particle. Therefore,  the state of the system after compression is equivalent to the state
 of the system, after insertion followed by the projective measurement.
 Similarly, during the removal process, the particle is delocalized due to tunneling, a factor that does not come into play 
 during
 expansion. Hence the extractable work during the removal process is less compared to that obtained during expansion.
 There is an element of lack of information due to the degeneracy (regarding the 
 particle's location)
 and tunneling which causes a difference in the amount of work.
 
The modeling of a quantum Szilard engine begins with the conversion of a single infinite potential well
to an infinite double well potential by introducing a barrier in the middle represented by
a delta function potential
of the form $\alpha\delta(0)$. Delta function potentials with positive values of $\alpha$ are regularly used in
the literature to represent thin
barriers~\cite{Moore-Book}, while negative values represent thin wells with attractive potentials~\cite{Vugalter2002}.
Consider a box of length $2a$ with rigid walls containing a single molecule
~\cite{Griffith-Book, Schiff-Book, Merzbacher-Book}. The slow insertion of the barrier 
corresponds to an increase in the 
value of $\alpha$ from zero to infinity, at the end of which the barrier is 
inserted completely.
When the barrier is introduced in the middle  completely and thereby
the system is converted to an infinite double well potential,
the even energy levels, each of which have a node at the origin, remain static
while the odd energy levels shift
upwards and overlap with the immediately next even energy levels~\cite{Belloni2014}. 
In practice, one can think of
a delta potential growing in strength from zero to a finite height $X_\infty$, that is large enough 
to prevent any tunneling through the barrier.
This can be ensured if the tunneling time exceeds the time required for the completion of the thermodynamic processes.
In this situation,
although the previously adjacent pairs of energy levels
do not overlap, they come sufficiently close to each other and become almost degenerate. This is more
pronounced in the lower levels, which, as we show below, plays a significant role for the regime we consider. 
One should also note that by odd and even energy levels we mean odd and even numbered energy levels
and do not refer to the parity of the energy eigenfunctions.

The main motivation for the present work is to devise a quantum heat engine which will work exclusively 
on quantum
features~\cite{Alan2017, Uzdin2015, Klatzow2017, Levy2018} 
and which may not operate in a classical regime. The quantum behavior, in our case, results 
from the energy quantization due to the small size of the potential well we consider. The Szilard
engine converts information into useful
work and hence
measurement is needed 
to extract work~\cite{Kim2011, Moore-Book, Li2012, Kim2012, Cai2012, Bracken2014, Zhuang2014, Cruz2016, Bengtsson2017}.
Our analysis
shows an effective way of converting lack of information, emerging from degeneracy of energy levels, to useful
work without any measurement but using two different reservoirs. The amount of extractable work depends on the nature 
of the particle such as distinguishable particles, bosons or fermions \cite{Kim2011, Kim2012,Cai2012,Zhuang2014,
Cruz2014,Jeon2016,Jaramillo2016,Bengtsson2017}. The cycle we use is quite similar to the 
Stirling cycle. Quantum versions
of Stirling engines  have been studied in the recent past \cite{Saygin2001, Chaturvedi2013,Huang2014,Thomas2017}.
In this paper, we provide a model of a quantum Stirling engine whose efficiency approaches the Carnot value at
low temperature limit.  It is to be noted that in a modified version of Szilard engine,
Carnot efficiency can be achieved by erasing
the information (obtained from the measurement) using a  heat bath of lower temperature compared to the one attached
to the engine \cite{Lloyd1997,Cai2012}. 
We also discuss the case where more than one partitions are inserted with a greater number of particles.

The paper is organized as follows: In Sec. \ref{section2}, we give a brief description of the quantum model of
a Stirling-like
engine. Section \ref{section3} consists of a brief discussion that points out the distinguishing 
features of our engine compared to the 
more conventional one that is based on expansion and compression. 
Section \ref{section4} is devoted to the limiting cases and discusses the behavior in a low
 temperature (reversible) limit. In Sec. \ref{section5}, we discuss the amount of work extractable 
from distinguishable particles and indistinguishable particles (fermions and bosons).
We conclude the paper with discussion in Section \ref{section6}.
\section{Stirling-like cycle}
\label{section2}
\begin{figure}[H]
\centering
 \includegraphics[scale=0.5]{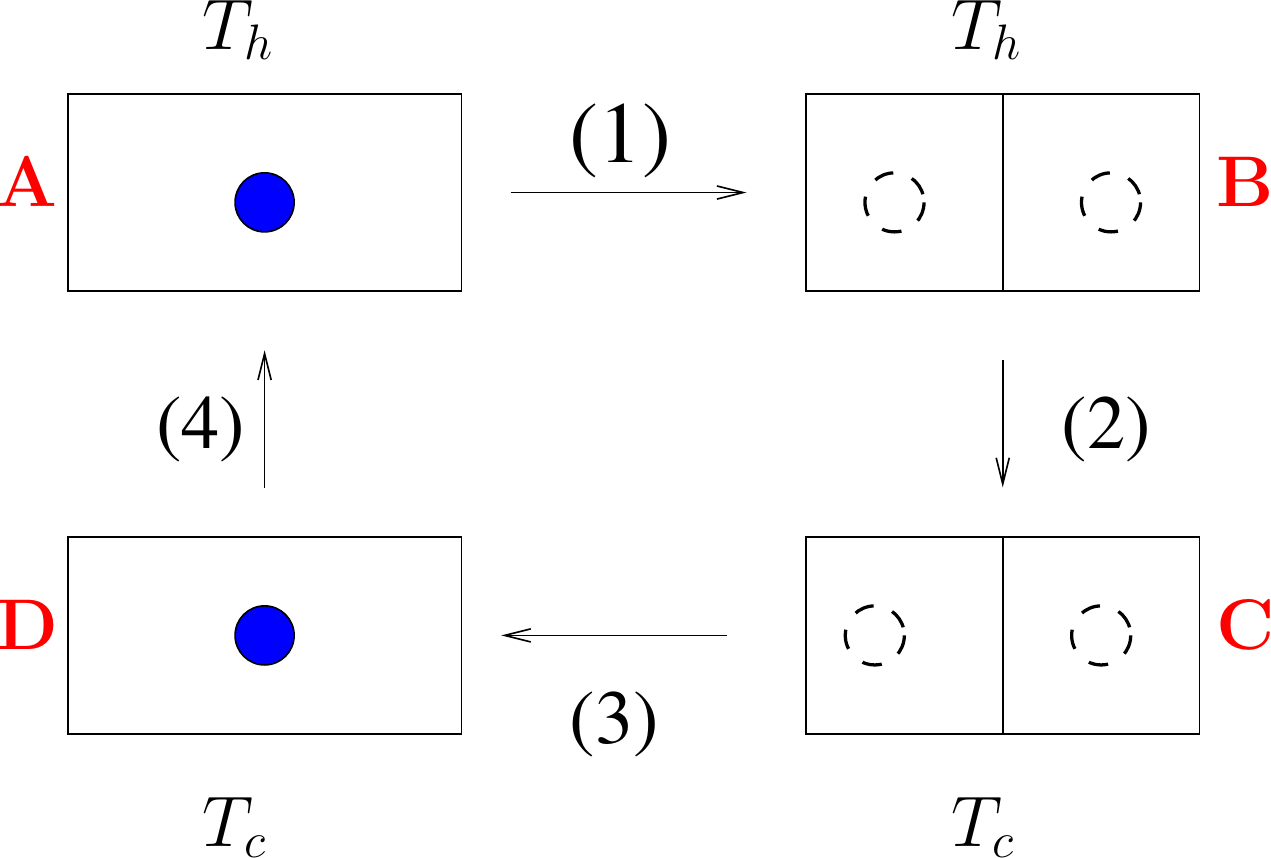}
 \caption{The four stages of the Stirling cycle. Stage 1 is isothermal insertion, stage 3 is isothermal removal and
 stages 2 and 4 are isochoric
 processes. The dashed circles in B and C signify ignorance about the location of the particle, which is 
 represented by solid circles in A and D.}
 \label{quantum_stirling_pic}
\end{figure}

A Stirling cycle~\cite{ Saygin2001,Valentin2011, Chaturvedi2013, Huang2014} consists of four stages, two
isothermal processes and two isochoric processes.
In the first stage, a barrier is inserted isothermally in the middle of the well such that
the working medium is in equilibrium with a hot bath at a temperature $T_h$ during the quasistatic insertion process.
In the second stage, the working medium undergoes isochoric heat exchange by connecting it with a heat
bath at a lower temperature $T_c$. Next, an isothermal removal of the barrier is effected by keeping the engine in equilibrium 
with the lower temperature bath at $T_c$. In the final stage, the engine is once again connected to 
the hot bath at the temperature $T_h$
and an isochoric heat absorption is carried out. The process is pictorially represented in Fig. \ref{quantum_stirling_pic}
Let us consider a particle of mass $m$ inside a one dimensional potential well of length $2a$,
the well being
at equilibrium with a bath of temperature $T_h$. The $n$th energy level of the one dimensional
potential well is given by
\begin{equation}
 E_{n}=\frac{n^2\pi^2\hbar^2}{2m(2a)^2}\;\;\;\;\;(n=1,2,3\cdots).
 \label{energyvalues}
\end{equation}
This can be used to calculate the  corresponding partition function of the system, given by
\begin{eqnarray}
 Z_A &=&\sum_{n=1}^{\infty} e^{-\frac{ E_{n}}{k_B T_h}}=\sum_{n=1}^{\infty} e^{-\frac{ n^2 \pi^2 \hbar^2}{2m (2a)^2 k_B T_h}}
 \label{partfnA}
\end{eqnarray}
where $k_B$ is the Boltzmann constant. The energy levels in Eq.~(\ref{energyvalues}) are evidently non-degenerate.

Suppose that a wall is inserted slowly in the middle of the box isothermally at this point.
As discussed earlier, this is achieved by increasing the value of $\alpha$,
in $\alpha\delta(0)$, from zero to infinity. For all the
subsequent analyses and discussions we consider the middle point as the origin of coordinates.
Immediately, the problem is then converted into an infinite double well potential. The
energy levels get reoriented as a result of this action. The energy levels corresponding to 
even values of $n$ remain unchanged while each energy level with odd value of $n$ shifts 
upwards and overlaps with the nearest neighboring even energy level
of the original single-well potential. This leads to a degeneracy in the energy
levels of this new setup (i.e. for the double-well ). We can thereby
express an arbitrary energy level of the partitioned one dimensional potential well as 
\begin{equation}
 E^\prime_{n}=\frac{(2n)^2\pi^2\hbar^2}{2m(2a)^2} \mathrm{, with}\; n=1,2,....
\end{equation}
Accordingly, the new partition function becomes
\begin{eqnarray}
 Z_B = \sum_{n=1}^{\infty} 2e^{-\frac{(2n)^2\pi^2\hbar^2}{2m (2a)^2 k_B T_h}}:=2Z_{a,T_h},
 \label{partfnB}
\end{eqnarray}
where $Z_{a,T_h}$ is the canonical partition function for a particle in potential well with length $a$ 
which is in thermal equilibrium with temperature $T_h$. The pre-factor 2 in 
Eq.~(\ref{partfnB}) arises because, due to the insertion of the barrier at the midpoint of the potential well,
the even energy levels of the original potential well become two-fold degenerate.
The internal energies $U_A$ and $U_B$ of the system can be calculated by
employing the respective partition functions $Z_A$ and $Z_B$ from Eqs
~(\ref{partfnA}) and~(\ref{partfnB}) as
 $U_{A/B}=-\partial \ln{ Z_{A/B}}/\partial \beta_h$, where $\beta_h=\frac{1}{k_BT_h}$.
The heat exchanged in the isothermal process (stage 1 in Fig.~\ref{quantum_stirling_pic}) of introducing the wall is thus,
\begin{equation}
 Q_{AB}=U_B-U_A+k_BT_h\ln{Z_B}-k_BT_h\ln{Z_A}.
\end{equation}
In the next step (stage 2 in Fig.~\ref{quantum_stirling_pic}), the system is connected to a heat bath at a lower
temperature $T_c$ after disconnecting it from the bath at temperature $T_h$. The energy levels
remain the same,
while the new partition function is given by
\begin{eqnarray}
 Z_C &=& \sum_{n=1}^{\infty} 2e^{-\frac{ E^\prime_{n}}{k_B T_c}}:=2Z_{a,T_c},
 \label{partfnC}
\end{eqnarray}
where $Z_{a,T_c}$ is the canonical partition function for a particle in a potential well with length $a$ 
which is in thermal equilibrium with a bath at temperature $T_c$.
The pre-factor in Eq.~(\ref{partfnC}) appears for the same reason as that for
Eq.~(\ref{partfnB}). 
The heat exchanged is now the difference of the average energies of the initial and the final 
configurations.
\begin{equation}
 Q_{BC}=U_C-U_B
 \label{qbc}
\end{equation}
with $U_C=-\partial \ln{ Z_C}/\partial \beta_c$ as the internal energy in the state C where
$\beta_c=\frac{1}{k_BT_c}$.
The wall is now removed slowly and isothermally (see stage 3 in Fig.~\ref{quantum_stirling_pic}), with the system
connected to the heat bath at temperature $T_c$.
The energy levels are once again restored to initial values given in Eq.~(\ref{energyvalues}).
while the partition function is now given by
\begin{eqnarray}
 Z_D =\sum_{n=1}^{\infty} e^{-\frac{ E_{n}}{k_B T_c}}
 =\sum_{n=1}^{\infty} e^{-\frac{ n^2 \pi^2 \hbar^2}{2m (2a)^2 k_B T_c}}.
\end{eqnarray}
If $U_D=-\partial \ln{ Z_D}/\partial \beta_c$ is the internal energy in the state D, the heat exchanged in the process is
\begin{equation}
 Q_{CD}=U_D-U_C+k_B T_c \ln{Z_D} -k_B T_c \ln{Z_C}
\end{equation}
In the final step (stage 4 in Fig.~\ref{quantum_stirling_pic}), the system is connected to the higher temperature
bath at $T_h$ once again. The energy levels
remain unchanged but the partition function changes to $Z_A$. The corresponding heat exchanged is given by
\begin{equation}
 Q_{DA}=U_A-U_D
 \label{qda}
\end{equation}
In the entire cyclic process, the total work done is, therefore
\begin{eqnarray}
 W&=&Q_{AB}+Q_{BC}+Q_{CD}+Q_{DA}\nonumber\\
 &=&k_B T_h\ln\frac{{Z_B}}{{Z_A}}-k_B T_c\ln\frac{{Z_C}}{{Z_D}}.
 \label{heated_sum}
\end{eqnarray}
Hence, the efficiency of the cycle is given by
\begin{equation}
 \eta=\frac{\textrm{total work done}}{\textrm{heat supplied}}=1+\frac{Q_{BC}+Q_{CD}}{Q_{DA}+Q_{AB}}
 \label{efficiency}
 \end{equation}
 It is to be noted that our engine represents an idealized case where the isothermal processes are done slowly 
 enough (compared to the tunneling time scales) to keep the system in equilibrium throughout the processes. We also consider
 the energy needed to couple and decouple the system with the baths is negligible.
 
The same methodology can be employed to investigate the cases where more than one barrier
are inserted inside the potential box. Consider the case of two barriers, inserted at distances 
$\frac{2a}{3}$ from the two walls of the box. Note that the third energy eigenstate, in the case of
no barrier, has nodes at the above two points. This implies that upon insertion 
of the two barriers, the third energy level remains unchanged while the first and the second energy 
levels shift to the third level. Similarly, all the energy levels in multiples of three remain 
unchanged on inserting the two barriers and the others shift accordingly. Thus an infinite triple potential 
well has energy levels that are triply degenerate. Hence, for the case of $N-1$ barriers, the original 
energy levels that are multiples of $N$ remain unchanged while the others shift and become degenerate with the 
former. This makes the energy levels of a potential well with $N-1$ barriers degenerate, with degeneracy
$N$.
\section{Comparison with the conventional scenario}
\label{section3}
\begin{figure}
\centering
 \includegraphics[scale=0.4]{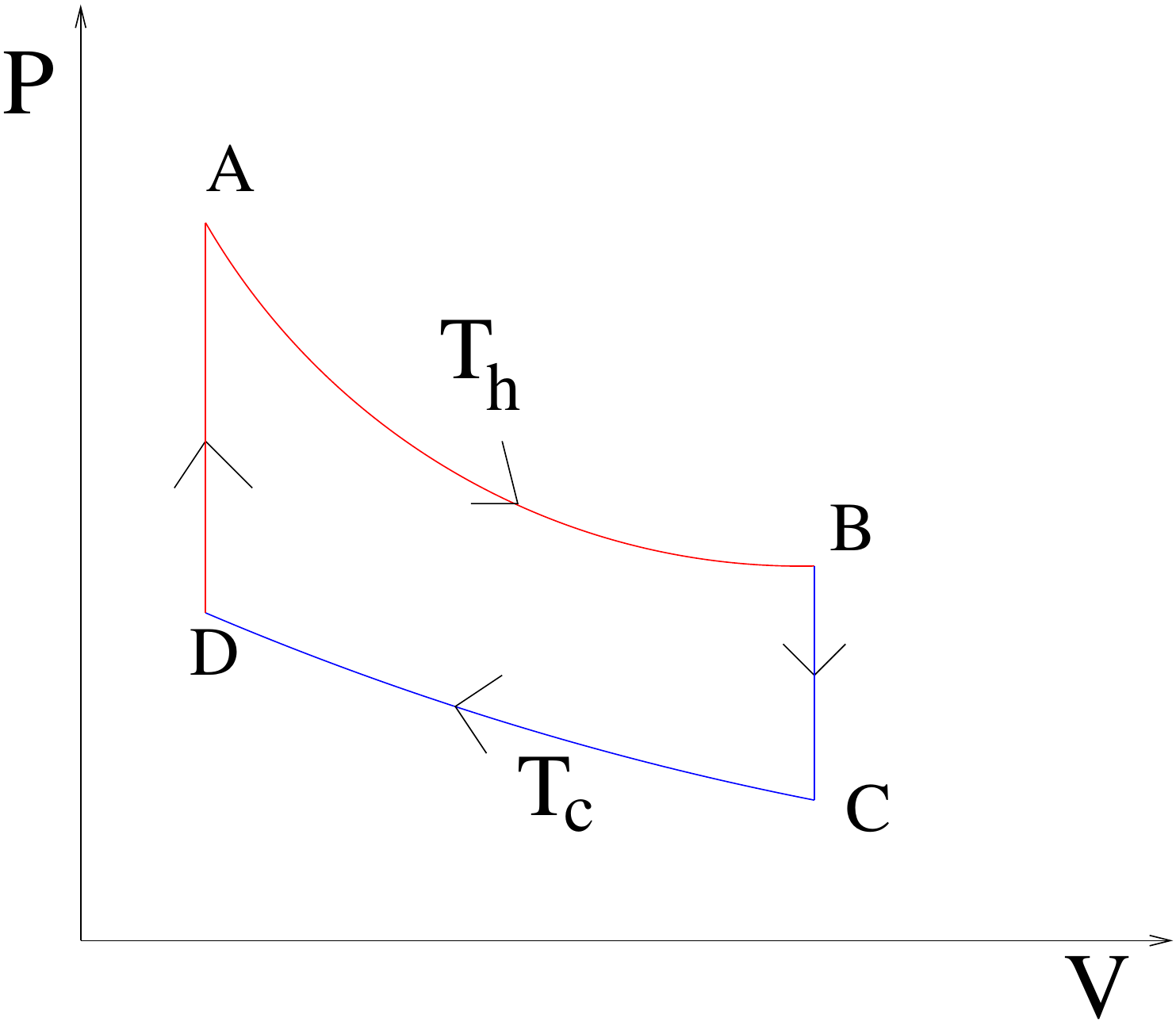}
 \caption{Pressure-volume ($P-V$) diagram for a classical Stirling cycle: AB and CD are the isothermal processes. BC and DA are
isochoric (constant volume) processes. The system is in contact with the hot bath
during DA and AB. The system is in contact with the cold bath during BC and
CD. 
Work is done only during isothermal branches.}
\label{pvdiagram}
\end{figure}

In a conventional Stirling engine, the isothermal expansion  is carried out by keeping the engine in 
equilibrium with a hot bath
 while the compression is carried out using the cold bath in contact, as given in Fig. \ref{pvdiagram}.
 On the other hand, in our cycle, the insertion is done when the system is in contact with the hot bath
 during the isothermal process $(1)$ in Fig.~\ref{quantum_stirling_pic}
 and isothermal removal of the barrier is assisted with the cold bath during the process $(3)$ in
 Fig.~\ref{quantum_stirling_pic}.
Consider the stage 1, discussed in Sec. \ref{section2}, a particle in the infinite potential well of length $2a$
and in equilibrium
with  a bath of temperature $T_h$.
The canonical partition function is $Z_A$. 
Now, consider a process in which we isothermally insert a barrier in the middle of the box.
The partition function at the end of the process
is $Z_B=2Z_{a,T_h}$ where $Z_{a,T_h}$ is partition function for particle trapped in box of length $a$. 
The factor 2 appears because of the degeneracy or in other words, due to the 
ignorance about the particle being in the left or right side of the box.
Therefore, the work done by the engine is the difference in free energies. 
$W_{\rm ins}=k_B T_h [\ln{2}+\ln{Z_{a,T_h}}-\ln{Z_A}]$. This work is less than that is
needed to compress the box from $2a$ to $a$. 
In the latter case, the amount of work needed is 
$W_{\rm com}=k_B T_h [\ln{Z_{a,T_h}}-\ln{Z_A}]$.
This is due to the fact that in the compression scenario,
the position of the particle is known whereas in the insertion scenario,
there is a lack of knowledge about the position of the particle.  
Similarly, the work extracted in the removal process is $W_{\rm rem}=k_B T_c[\ln{Z_D}-\ln{2}-\ln{Z_{a,T_c}}]$. 
In this paper, we  consider a cycle in which, we insert the barrier when the system is attached to
a hot bath of temperature $T_h$ and remove when the system is attached to a cold bath of temperature $T_c$. 
Therefore the net work done by the system (Eq. (\ref{heated_sum})) in our model is given by
\begin{eqnarray}
W &=& W_{\rm ins}+W_{\rm rem}\nonumber\\
  &=& k_B (T_h-T_c)\ln{2}+  k_B T_h\ln{\left(\frac{Z_{a,T_h}}{{Z_A}}\right)}\nonumber\\
  &&-k_B T_c\ln{\left(\frac{Z_{a,T_c}}{{Z_D}}\right)}.\nonumber\\
  .
 \label{Work_con}
\end{eqnarray}
The second term of the right hand side of Eq. (\ref{Work_con}) is equivalent to the work done
 during an isothermal compression  from a box of length $2a$ to $a$ at the higher temperature $T_h$. On the other hand,
 the third term is equal to the work done during an isothermal expansion of a box of length $a$ to $2a$ at the lower 
 temperature $T_c$.
 In a certain limiting case, we show that the work done by the engine
is $k_B(T_h-T_c) \ln{2}$ , i.e., the second and third terms in Eq. (\ref{Work_con}) cancel each other. 
The appearance of $\ln{2}$ is due to the change in the entropy of the working fluid during insertion and removal of the barrier because of the degenerate energy levels (causing
ignorance about the position of the particle).
\section{Low temperature Limit}
\label{section4}
In this section, we discuss the  extractable work  from our model in the low temperature limit.
In this limit
~\cite{Kim2011,Kim2012,Cai2012}, 
the system works in an almost reversible
manner at near Carnot efficiency. 

%\subsection{ Low temperature limit}
Let us consider a box with length $2a$ such that $\pi^2\hbar^2/2m(2a)^2>>k_BT_{h}$, where $T_h>T_c$.
It can be seen that the above condition holds good for lower values of temperatures
$T_h$ as well as small values of $a$. We will refer to this case as the low temperature
limit in all subsequent discussions.
In the low temperature limit, at the beginning of stage 1 (point A in Fig. \ref{quantum_stirling_pic}), the occupational probability in the ground state
is close to unity and the entropy of the system approaches zero. 
When the partition is inserted (during stage 1), the ground state of the double well becomes 
doubly degenerate with occupational probability $1/2$ 
for each state and hence the entropy becomes $k_B\ln{2}$. Therefore, the
total heat absorbed by the system from the hot bath becomes $k_BT_h \ln{2}$.
Similarly, when the wall is removed, the heat exchanged between the system and the cold bath is $-k_BT_c \ln{2}$. 
The second and third terms in Eq. (\ref{Work_con}) cancel each other in the low temperature limit.
Therefore, the work done and the efficiency in this case
become respectively
\begin{equation}
 W\approx k_B(T_h-T_c)\ln{2},\;\;\; \eta \approx 1-\frac{T_c}{T_h}.
\end{equation}
\begin{figure}
\centering
\includegraphics[scale=1]{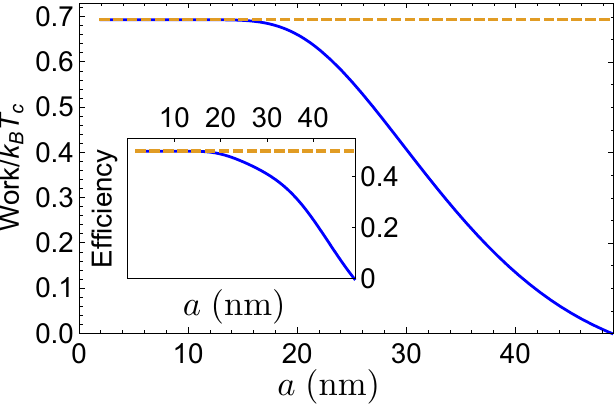}
 \caption{Plot of $\rm W/k_BT_c$, with $W$ of Eq. (\ref{heated_sum}), vs $a$ (in nanometers), the width 
 of each well of a double well potential. The horizontal dotted line represents the low temperature limiting case
  (with $a\rightarrow 0$).
 Inset shows the plot of efficiency Eq.(\ref{efficiency}) vs $a$. The horizontal line for the inset represents the Carnot
efficiency ($1-\frac{T_c}{T_h}$) obtained from the  low temperature limit.
 Here, we have taken $m=9.11 \times 10^{-31}$ kg, $T_h=2K$ and $T_c=1K$.}
\label{plot_W_eta}
\end{figure}
 The system can, therefore, nearly attain Carnot efficiency and hence it is
 almost reversible. 
 The dimensionless work ($W/k_B T_c$) 
 and the efficiency are plotted with the length $a$ of the box
 in Fig. \ref{plot_W_eta} and the corresponding values of these two quantities, for the low temperature limit,
 are also depicted. 
 It is to be noted that, for a two-level system at temperature $T$ with energy-level spacing
 $\omega$, the canonical heat
capacity  can be written as  $\frac{\partial U}{\partial T}|_{\omega}=(\omega^2/k_B T^2)\exp{(\omega/k_B T)}/[1+\exp{(\omega/k_B T)}]^2$,
where $U$ is the mean energy of the system~\cite{Johal2008, Johal2010, Levy2018}.
Therefore, for $\omega>>k_B T$, the heat capacity goes to zero. Analogously, one can see that in
the low temperature limit of the particle in a box, 
the heat capacity vanishes
and hence the heat exchanged to lower or raise the temperature during stage 2 [$Q_{BC}$ given in Eq. (\ref{qbc})] 
or stage 4 [$Q_{DA}$ given in Eq. (\ref{qda})] respectively also vanish.

It will be interesting to contrast the work extracted for our cycle in the low temperature limit with the conventional
compression-expansion based Stirling cycle. In the low temperature limit, the ground state population is close to unity
owing to the large energy-spacing between the first two levels compared to
thermal fluctuations. During the process of slow insertion of the barrier in the middle of the box,
the ground state of the system moves towards the first excited state, which remains stationary. After a 
certain time during the process, the two levels come close
enough to allow considerable transitions between them, resulting in decreased population in the ground state.
The changes in populations can be associated with the
heat exchanged between the system and the bath. At the end of the process, when the barrier is completely inserted,
the two energy levels overlap and become 
equally populated i.e the probability of occupancy of each is $\frac{1}{2}$. Hence the contribution of heat during insertion
is $k_B T_h \ln{2}$. On the other hand, if we 
analyze the isothermal compression, the width of the well is reduced and consequently, the gap between the two energy
levels increases. This further restricts the 
transitions between the ground state and the first excited state. Therefore the heat exchanged between the system and
the bath is close to zero. The sole contribution 
to the increase in the internal energy of the system comes from the work done on the system. This explains the 
difference in the amounts of work needed in the insertion
and the compression processes. Similar arguments can be made in the removal and expansion processes where the 
difference is $-k_B T_c \ln{2}$. In our cycle, for the 
low temperature limit, we exploit these heat exchange amounts with the baths.

%\subsection{Classical limit}
It is worthwhile to note here the  behavior of the Stirling-like cycle in the classical limit.
The classical limit
is obtained from a large width of the potential box (i.e., $a\rightarrow \infty$), where the particle behaves like
a free particle. 
In case of a potential box, the energy difference between the two adjacent levels (nth and (n+1)th)
is $(2n+1)\pi^2\hbar^2/2m(2a)^2$.
During the insertion of the barrier, the odd numbered energy levels approach the next even numbered levels.
When the
barrier is fully inserted, each energy level will be doubly degenerate. Therefore the gaps between the adjacent
energy levels are responsible for the work.
When $a\rightarrow \infty$, the energy gaps go to zero and there will be a continuum of energy levels.
Therefore the work required to insert or remove the barrier goes to zero, as the particle becomes a free particle in that limit.
Note that the classical limit has been explored earlier in the context of the Szilard engine in~\cite{Li2012}. 

\subsection{Refrigerator in low temperature limit}
A modified version of our cycle can be used as a quantum refrigerator where heat is transferred
from the cold to
the hot reservoir by doing work on the system \cite{
Gemmer-Book,Kosloff2000,Rezek2009,Allahverdyan2010,Chen2012,Huang2014,Thomas2017}.
In this model, the  isothermal insertion
is achieved by keeping the system in contact with the cold bath at temperature $T_c'$. In the 
low temperature limit, the heat absorbed by the refrigerator from the cold bath is $k_B T_c' \ln{2}$.
The system then undergoes an isochoric process by 
attaching it to a hot bath at temperature $T_h'$ ($T_h'>T_c'$).  Again, the isothermal removal of the barrier 
is carried out by attaching the system to the hot bath. 
To complete the cycle, the system is attached to the cold bath and 
another isochoric process is carried out. The coefficient of performance \cite{Callen-Book} of such a cycle
can approach the Carnot value  $T_c'/(T_h'-T_c')$ in the low temperature limit ($\pi^2\hbar^2/2m(2a)^2>>k_BT_{h}$). 
In this scenario, the first term in Eq. (\ref{Work_con})
becomes $k_B(T_c'-T_h')\ln{2}$, a negative quantity.  Correspondingly, the sum of the second and third terms
can be positive but negligible compared to
$k_B(T_h'-T_c')\ln{2}$ in the low temperature limit, resulting in the total system working as a refrigerator. 
But for a sufficiently larger value of $a$
and before reaching the classical limit, the sum of the second and third terms can be larger than $k_B(T_h'-T_c')\ln{2}$,
resulting in the system performing like a heat engine.
\section{Engine with distinguishable and indistinguishable particles}
\label{section5}
We have looked at the behavior of the work extracted from the Stirling engine at low temperature and classical limits.
In this section, we explore the properties of the working fluid manifesting as the amount of extractable work in the
low temperature limit.
Let us consider two fermions and two bosons in the low temperature limiting case with a single partition.
Before inserting the partition, the  system  is in the ground state (or at least highly
probable to be in the ground state) because of low temperature. The ground states of the system  for bosons ($\Psi_{B}$) and
fermions ($\Psi_{F}$) are given as
\begin{equation}
 \Psi_{B/F}=\frac{1}{\sqrt{2}}[\psi_{n_1}(x_1)\psi_{n_2}(x_2)\pm\psi_{n_2}(x_1)\psi_{n_1}(x_2)],
\end{equation}
where $\psi_{n_1}$ and $\psi_{n_2}$ represent the wavefunctions corresponding to the  $n_1^{th}$ and $n_2^{th}$ energy eigenstates.
The ground state for the case of fermion takes the values $n_1=1$ and $n_2=2$. On the other hand for bosons, both of
the particles can be in the same state and hence it can take $n_1=n_2=1$. 
Upon the insertion of the wall, the ground state becomes doubly degenerate. Moreover, the number of ways of
arranging the different classes of particles is also different. 

\begin{figure}[H]
\centering
 \includegraphics[scale=0.5]{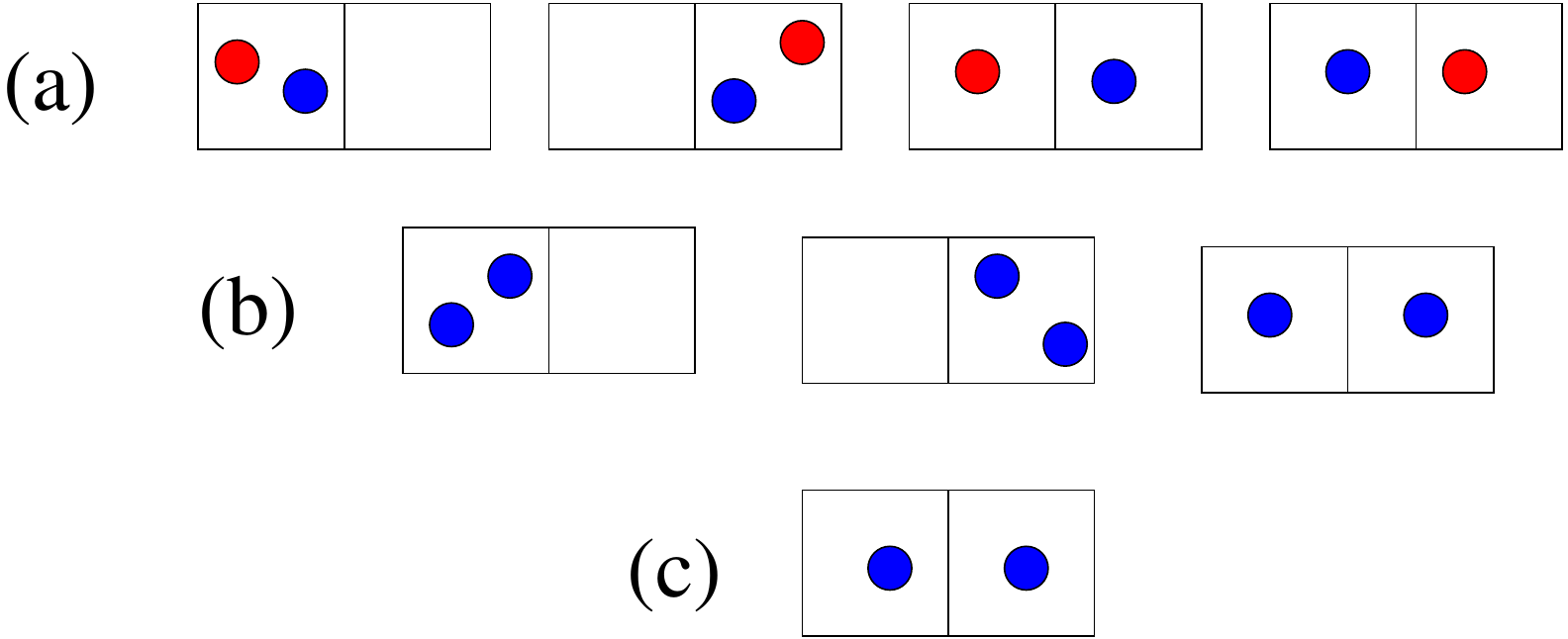}
 \caption{Particle statistics after inserting the barrier: (a) Distinguishable particles, (b) bosons and (c) fermions.}
 \label{particle_stat_fig}
\end{figure}

It is interesting to study the quantity of work extracted from the engine in the case of different classes
of particles and the effects of increasing the number of partitions in the potential well. We would like 
to clarify here that the present analysis relates only to the limiting case (ie., the low temperature regime). 
Suppose after keeping our potential 
box with two distinguishable particles in the ground state, in equilibrium with a bath at temperature $T_h$, 
we insert a single partition in the middle. From the previous discussions, we know that the energy levels are doubly 
degenerate. Thus 
two particles can occupy the two states of the lowest energy level, one in each state, in two possible ways. Again, two particles
can be in the same state in two different ways (see Fig.~\ref{particle_stat_fig}). Each of these possibilities comes with a probability $\frac{1}{4}$.
Thus the entropy of the system is $2\ln{2}$ and the heat absorbed from the hot reservoir upon isothermal insertion of 
the wall in the middle is $2k_B T_h\ln{2}$. Now if the system is connected to a heat bath at a lower temperature
$T_c$ and the wall is removed isothermally, by the previous argument, the heat released is $2k_B T_c\ln{2}$. Thus
the work done by the system is $2k_B (T_h-T_c)\ln{2}$.
The situation becomes even more exciting in the case of two fermions in the ground state. There is only one 
configuration in which two fermions can be arranged in the two states of the same energy level after the barrier
is inserted. Hence, the changes in entropy during insertion and removal processes are zero and consequently no work can 
be extracted from the engine in the case of two fermions in the ground state. This must be contrasted with the 
case in which the potential well contains two bosons in the ground state and the work extracted out of a Stirling-like
cycle performed on it. In the ground state, two bosons can have three possible configurations, hence the 
change in entropy upon insertion or removal of the barrier is $k_B\ln{3}$. Thus the work that can be extracted out
of the engine is $k_B (T_h-T_c)\ln{3}$.

Let us now examine the case in which we have a potential well with two distinguishable particles and insert two 
partitions, isothermally, at $-\frac{a}{3}$ and $\frac{a}{3}$. Upon insertion at a temperature $T_h$, the change
in entropy is $2\ln{3}$ and the heat exchanged is $2k_BT_h\ln{3}$ as each energy level acquires a three-fold degeneracy.
Similarly the heat exchanged during the 
isothermal removal of the walls at a temperature $T_c$ is $2k_BT_c\ln{3}$. The amount of work that can now be 
extracted from the engine is $2k_B(T_h-T_c)\ln{3}$. Fermions, however, can occupy the three states of the ground
level in three possible ways only and hence the work extracted can be only $k_B(T_h-T_c)\ln{3}$. Bosons, on the 
other hand can occupy these states in six possible ways and therefore the work done is $k_B(T_h-T_c)\ln{6}$. In
general, upon inserting $g$ partitions in a box with $n$ particles, there are  $(g+1)^n$ possible ways
to arrange distinguishable particles in the degenerate ground states, while bosons and fermions can be arranged 
in $(n+g)!/(n!g!)$ and $(g+1)!/(n!(g+1-n)!)$ numbers of different ways,
respectively.
The entire discussion is summarized in Table~\ref{table1}. The magnitudes of work done by the engine for three
particles of different classes is also summarized in Table~\ref{table2}. It is observed that for
a given number of particles and partitions,
the maximum work is extracted from distinguishable particles followed by bosons and then by fermions.

\begin{center}
\begin{table}
\caption{Comparison of work  for the case of two particles. 
\label{table1}}
\begin{center}
\begin{tabular}{llll}
\hline
\hline
 Particles~~~&Barriers~~~~ & Work \\
\hline 
Distinguishable~~~~~ &1& $k_B(T_h-T_c)\ln{2^{2 }}$ \\
\hline 
 Bosons ~~~~~  &1 & $ k_B(T_h-T_c)\ln{3}$ \\
\hline 
 Fermions~~~~~   &1 & $ 0$  \\
\hline 
Distinguishable~~~~~ &2 & $k_B(T_h-T_c) \ln{3^{2}}$ \\
\hline 
 Bosons ~~~~~ &2& $ k_B(T_h-T_c)\ln{6}$ \\
\hline
 Fermions~~~~~ &2 & $k_B(T_h-T_c)\ln{3}$ \\
\hline 
%%%%%%%%%%%%%%%%%%%%%%%%%%%%%%%% 
%%%%%%%%%%%%%%%%%%%%%%%%%%%%%%%%%%%%
\end{tabular} 
\end{center}
\end{table}
%%%%%%%%%%%%%%%%%%%%%%%%%%%%%%%%%%%%
\end{center}
\begin{center}
\begin{table}
\caption{Comparison for the case of three particles. 
\label{table2}}
\begin{center}
\begin{tabular}{llll}
\hline
\hline
 Particles~~~&Barriers~~~~ & Work \\
\hline 
Distinguishable~~~~~ &1& $ k_B(T_h-T_c) \ln{2^{3}}$ \\
\hline 
 Bosons ~~~~~  &1 & $k_B(T_h-T_c) \ln{2^{2}}$ \\
\hline
 Fermions~~~~~   &1 & $k_B(T_h-T_c) \ln{2}$  \\
\hline 
 Distinguishable~~~~~ &2 & $k_B(T_h-T_c) \ln{3^{3}}$ \\
\hline 
 Bosons ~~~~~ &2& $k_B(T_h-T_c) \ln{10}$ \\
\hline
 Fermions~~~~~ &2 & 0 \\
\hline 

%%%%%%%%%%%%%%%%%%%%%%%%%%%%%%%% 
%%%%%%%%%%%%%%%%%%%%%%%%%%%%%%%%%%%%
\end{tabular} 
\end{center}
\end{table}
%%%%%%%%%%%%%%%%%%%%%%%%%%%%%%%%%%%%
\end{center}
\section{Discussions and conclusion}
\label{section6}
We considered a Stirling-like cycle which uses quantized energy levels to extract work.
The lack of knowledge of the particle's position can be effectively converted into useful work without involving
measurement to locate the particle.
Our engine operates exclusively using quantum features and does not work in the classical limit where the  width
of the box is large. 
In the low temperature limit our engine approaches the Carnot efficiency.
The work obtained from the engine 
depends upon the number of partitions and the number particles as well as the spin-statistics nature of the particles. 
The extractable work from distinguishable particles,
fermions, and bosons is compared.

\begin{figure}[t]
\centering
\includegraphics[scale=.9]{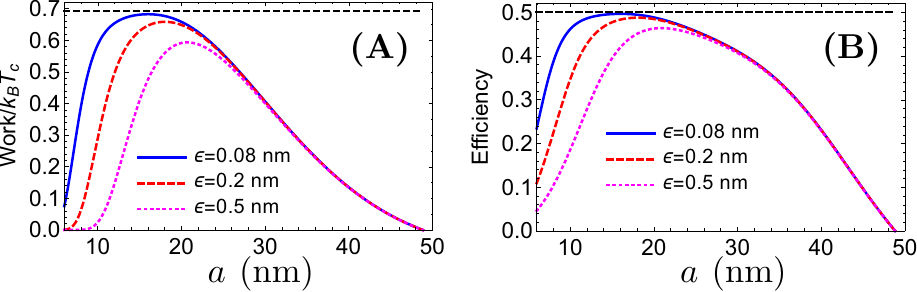}
 \caption{(a) Plot of $\rm {work}/k_BT_c$  vs the $a$ which is the  half width  of the total potential well,
 for different values of $\epsilon$ in nanometers. The horizontal line represents the low temperature limiting
 case value $(T_h-T_c) \ln{2}/T_c$.
 (b) The plot shows the behavior of efficiency vs $a$ for different values $\epsilon$. The horizontal line
 represents the Carnot efficiency ($1-\frac{T_c}{T_h}$) obtained from the  limiting case.
 Here, we have taken $m=9.11 \times 10^{-31}$ kg, $T_h=2K$ and $T_c=1K$.}
\label{plot_W_eta_asy}
\end{figure}

It is worth noting that we have discussed the effects of inserting one or more partitions, on the energy levels 
of a potential well at particular points. To start off, we note that all the wave functions 
corresponding to even numbered energy levels of an infinite single 
potential well have nodes at the origin. Hence, inserting a partition at the origin leaves them unchanged. Similarly,
all energy levels with multiples of three have nodes at $-\frac{a}{3}$ and $\frac{a}{3}$. Thus to leave these energy
levels unchanged, it is required to insert the barrier at these precise points. The same argument holds for energy
levels with
multiples of $N$, an arbitrary integer. Considering more practical situations, it is useful to explore the effects of
inserting one
or more partitions at some other points. All the energy levels would then shift resulting in different amounts of work.
Particularly, if a single partition is inserted $\epsilon$ distance away from the origin, say to the left, then 
the original first and second energy levels before insertion, do not completely merge but remain very close 
to each other. The width of the right and left  wells are now $a+\epsilon$ and $a-\epsilon$, respectively.
Hence we get nearly degenerate levels for small $\epsilon$. An effect of degeneracy is that the additional 
term $\ln{2}$ in the work extracted. The near degeneracy ensures a value that is close to $\ln{2}$.
The efficiency of such an engine, in the low temperature limit, is thus close to the Carnot value as discussed earlier.
However, for large $\epsilon$,
the shifts do not bring the energy levels close enough, resulting in no such term. The work and efficiency of such an engine
would be significantly lower. For different values of $\epsilon$, the work and efficiency are plotted versus the half-width
of the total potential well in Fig. \ref{plot_W_eta_asy}. A study  in 
which the barrier is inserted asymmetrically and adiabatically can be found in~\cite{Joakim2019}.
It is to be noted that in all our analysis, we restrict the length of the box to be much greater than
the Compton wavelength and hence our analysis is completely non-relativistic \cite{Gil1996}.

A future direction includes modeling a heat engine with finite-time processes with a finite barrier. For 
practical purposes, such cycle may be of interest due to finite power.
Our model can be applied to any other potential where the insertion of the barrier leads to degenerate
  or nearly degenerate eigenstates. Apart from an infinite square-well potential, one of the alternatives
  is a harmonic potential.
One can also take
different forms of potentials with interacting particles \cite{Bengtsson2017}.  A micrometer-sized Stirling
engine has already been realized with a single colloidal particle \cite{Valentin2011}.
In~\cite{Alan2017} a double-well infinite potential with or without a delta-function barrier has been mimicked by a
laser-cooled trapped ion in a combined potential of a Paul ion trap and a sinusoidal potential of an optical lattice. The 
potential has been used to implement an Otto cycle, enabled by energy quantization and operating by adiabatic insertion and 
removal of the barrier.
A possible candidate to realize our model of the quantum Stirling engine
involves superconducting flux qubits where the symmetric potential can be controlled at very low temperatures
\cite{Chiarello2007,Poletto2009} with well defined heat baths and the possibility
of measurement of heat power \cite{Jukka2018}.

\section*{Acknowledgements}
 The authors acknowledge Dmitry S. Golubev for useful discussions.
 %\bibliography{szilard.bib}
%\end{document}
 
%
\end{document}